\documentclass[doublecol,amsmath,amssymb,revision]{epl2}
\usepackage{amsmath}
\usepackage{cases}
\usepackage{amsfonts}
\usepackage{color}
\usepackage{graphicx}
\usepackage{dcolumn}
\usepackage{bm}
\usepackage{amssymb}
\usepackage{array}
\usepackage{longtable}
\usepackage{hhline}

\begin{document}

\title{Exact eigenvalue spectrum of a class of fractal scale-free networks}
\shorttitle{Exact eigenvalue spectrum of a class of fractal scale-free networks}

\author{Zhongzhi Zhang\inst{1,2} \footnote{\email{zhangzz@fudan.edu.cn}} \and Zhengyi Hu\inst{1,2} \and Yibin Sheng\inst{3} \and Guanrong Chen \inst{4}}
\shortauthor{Zhongzhi Zhang, Zhengyi Hu, Yibin Sheng, and Guanrong Chen}

 \institute{
  \inst{1} School of Computer Science, Fudan University, Shanghai 200433, China\\
  \inst{2} Shanghai Key Lab of Intelligent Information Processing, Fudan University, Shanghai 200433, China\\
  \inst{3} School of Mathematical Sciences, Fudan University,
Shanghai 200433, China\\
  \inst{4} Department of Electronic Engineering, City University of Hong Kong, Hong Kong SAR, China}

\date{\today}

\begin{abstract}{
The eigenvalue spectrum of the transition matrix of a network encodes important information about its structural and dynamical properties. We study the transition matrix of a family of fractal scale-free networks and analytically determine all the eigenvalues and their degeneracies. We then use these eigenvalues to evaluate the closed-form solution to the eigentime for random walks on the networks under consideration. Through the connection between the spectrum of transition matrix and the number of spanning trees, we corroborate the obtained eigenvalues and their multiplicities.}
\end{abstract}



 \maketitle

\section{Introduction}

Complex networks have emerged as a universal tool for studying complex systems in many different fields, such as physics, chemistry, biology, and computer science~\cite{AlBa02,DoMe02,Ne10}. A fundamental issue in network science is to understand diverse dynamical processes defined on different networks~\cite{Ne03,BoLaMoChHw06,DoGoMe08}. A paradigmatic dynamical process that has attracted increasing research interest is random walks~\cite{NoRi04,SoRebe05,Bobe05,CoBeTeVoKl07,GaSoHaMa07,KiCaHaAr08,TeBeVo09,FrFr09,RoHa11,TeBeVo11,ZhAlHoZhCh11}, which have found a broad range of applications in various areas of science and engineering~\cite{We94,Hu95,Re01,BeLoMoVo11}.

As is well known, random walks on a network can be described by a transition matrix and various interesting quantities about random walks are closely related to the spectrum of the transition matrix~\cite{ZhAlHoZhCh11,Lo96,AlFi99,LeLo02}. For example, the mean first-passage time (MFPT) from one node to another on a network can be expressed in terms of the eigenvalues and orthonormalized eigenvectors of its transition matrix~\cite{ZhAlHoZhCh11,Lo96,AlFi99}; the sum of reciprocals of one minus each eigenvalue, other than the eigenvalue $1$, determines the eigentime identity~\cite{AlFi99}, sometime referred to as Kemeny's constant~\cite{LeLo02}; the second largest eigenvalue, together with the smallest eigenvalue, determines the mix time of random walks on the network~\cite{Lo96,AlFi99}. Thus, it is of theoretical and practical significance to calculate and analyze the spectra of the transition matrices of networks. However, little attention has been paid to this relevant issue, the main reason for which is that the computational complexity for computing the eigenvalue spectrum of a general network is very high~\cite{Bi93}.

In this paper, we present a theoretical study of the transition matrix for unbiased random walks on a family of iterative networks~\cite{SoHaMa06,RoHaAv07,RoAv07}, which have remarkable scale-free~\cite{BaAl99} and fractal~\cite{SoHaMa05} phenomena that constitute our fundamental understanding of real-life systems. Making use of the real-space decimation technique~\cite{DoAlBeKa83,CoKa92}, we will derive an exact recursion expression for the eigenvalues at every two consecutive iterations, which is obtained based on the structure of the studied networks. We then proceed to provide the complete spectrum of eigenvalues and their corresponding degeneracies. Using these eigenvalues, we further derive an explicit formula for the eigentime identity. Finally, we prove the validity of the computation for eigenvalues based on the relationship between the number of spanning trees and the eigenvlaues of the transition matrix of a network.


\begin{figure}
\begin{center}
\includegraphics[width=0.7\linewidth]{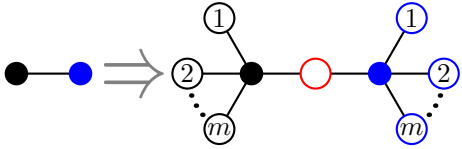}
\caption{ (Color online) Construction of the
networks. One can obtain the next generation of the network family through replacing each edge of the present generation by the clusters on the right-hand side of the arrow.}
\label{cons}
\end{center}
\end{figure}

\section{Model and properties\label{model}}

\begin{figure}
\begin{center}
\includegraphics[width=0.95\linewidth]{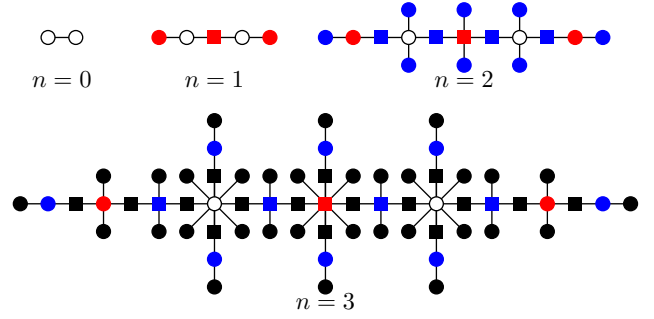}
\end{center}
\caption[kurzform]{ (Color online) Growth process for a
special case of network corresponding to $m=1$.} \label{network}
\end{figure}

The considered fractal scale-free networks are constructed in an iterative way~\cite{SoHaMa06,RoHaAv07,RoAv07}. Let $F_n$ ($n\geq 0$) denote the networks after $n$ iterations. For $n=0$, $F_0$ consists of an edge connecting two nodes. For $n\geq 1$, $F_n$ is obtained from $F_{n-1}$ by performing the following operations on every existing edge in $F_{n-1}$ as shown in
Fig.~\ref{cons}: replace the edge by a path of 2 links long, with both endpoints of the path being the identical endpoints of the original edge, then for each endpoint of the path create $m$ (a positive integer) new nodes and attach them to the endpoint.
Figure~\ref{network} illustrates the first several iterative construction steps for a particular case of $m=1$, whose dynamics of random walks were studied in~\cite{ZhGuXiQiZh09}.

The deterministic construction of the networks allows for treating exactly their relevant properties. At each generation $n_i$ ($n_i\geq 1$), the number of newly introduced nodes is $V_{n_i}=(2m+1)(2m+2)^{n_i-1}$. Then, the network size (number of nodes) $N_n$ of $F_n$ is given by
\begin{equation}\label{Nn}
N_n=\sum_{n_i=0}^{n}V_{n_i}=(2m+2)^{n}+1.
\end{equation}
And the total number of edges is
\begin{equation}\label{En}
E_n=V_n-1=(2m+2)^{n}\,.
\end{equation}
Let $d_i(n)$ be the degree of node $i$ in $F_n$ that entered the networks at generation $n_i$ ($n_i\geq 0$). Then $d_i(n+1)=(m+1)\,d_i(n)$, that is, after each new iteration the degree of node $i$ increases by $m$ times.

These resultant networks display the following interesting structural features. They are in power law with the degree distribution exponent being $\gamma=1+\ln (2m+2)/ \ln (m+1)$~\cite{RoHaAv07,ZhZhChGu08,ZhLiMa11}. In addition,  after the evolution of one generation, the number of nodes increases by a factor $f_N=2m+2$, see Eq.~(\ref{Nn});  the diameter grows by a factor of  $f_D=2$~\cite{RoHaAv07}, and the MFPT between two previously existing nodes increases by a factor of $f_M=4m+4$~\cite{RoHaAv07,ZhLiMa11}. Thus, the networks are fractal with a fractal dimension $f_B=\ln f_N/ \ln f_D=  \ln (2m+2)/ \ln 2$~\cite{SoHaMa06,GaSoHaMa07,ZhZhChGu08}; their random-walk
dimension is $f_w=\ln f_M/ \ln f_D=\ln(4m+4)/\ln 2$, and their spectral dimension is $f_s=2f_B/f_w=2\ln (2m+2)/\ln(4m+4)$. Finally, they are ``large-world'' with their diameter and average distance growing in a power of the network size~\cite{RoHaAv07,ZhZhChGu08,ZhLiMa11}.

\section{Spectrum of transition matrix}

After introducing the model and some properties of the networks, in this section we study the eigenvalue spectrum of the transition matrix for the network family, as well as their degeneracies.

\subsection{Eigenvalue spectrum}

As is well known, the structure of $F_{n}$ is encoded in its adjacency matrix $A_n$, whose entries $A_n(i,j)$ are defined by $A_n(i,j)=1$ if nodes $i$ and $j$ are adjacent in $F_{n}$, or $A_n(i,j)=0$ otherwise. Then, the transition matrix of $F_{n}$, denoted by $T_n$, is defined as $T_n=D_n^{-1}A_n$, where $D_n$ is the diagonal degree matrix of $F_{n}$ with its $i$th diagonal entry being $d_i(n)$. Thus, the element of $T_n$ is $T_n (i,j)= A_n(i,j)/d_i(n)$, which represents the jumping probability of the discrete-time unbiased random walks~\cite{NoRi04} for a particle going from node $i$ to node $j$.

We now consider the eigenvalue spectrum of $T_n$. Since $T_n$ is asymmetric, we introduce the following matrix:
\begin{equation}\label{Mat03}
P_n=D_n^{-\frac{1}{2}} A_n D_n^{-\frac{1}{2}}=D_n ^{\frac{1}{2}}T_n D_n^{-\frac{1}{2}}.
\end{equation}
Obviously, $P_n$ is real and symmetric and has the same set of eigenvalues as $T_n$.
It is easy to verify that the entry of $P_n$ is $P_n(i,j)=\frac{A_n(i,j)}{\sqrt{d_i(n)}\sqrt{d_j(n)}}$. Next, we apply the decimation method~\cite{DoAlBeKa83,CoKa92} to determine the eigenvalues and their multiplicities of $P_n$. The decimation approach is universal and has been used to compute the Laplacian spectra of Vicsek fractals~\cite{BlJuKoFe03,BlFeJuKo04,ZhWjZhZhGuWa10} and their extensions~\cite{JuvoBe11}.

We now address the eigenvalue problem for matrix $P_{n+1}$.
Let $\alpha$ represent the set of nodes belonging to $F_{n}$, and $\beta$ the set of nodes generated at $(n+1)$th iteration. By definition, $P_{n+1}$ has a block form
\begin{equation}\label{T2}
P_{n+1}=\left[\begin{array}{cccc}
P_{\alpha,\alpha} & P_{\alpha, \beta} \\
P_{\beta,\alpha} & P_{\beta, \beta}
\end{array}
\right]
=\left[\begin{array}{cccc}
0 & P_{\alpha, \beta} \\
P_{\beta,\alpha} & 0
\end{array}
\right],
\end{equation}
where we have used the facts that $P_{\alpha,\alpha}$ is the zero matrix of order $N_n\times N_n$ since there are no transitions between any pair of original nodes, and that $P_{\beta, \beta}$ is the zero matrix of order $(N_{n+1}-N_n)\times (N_{n+1}-N_n)$.

Suppose $\lambda_{i}(n+1)$ is an eigenvalue of $P_{n+1}$, and $u=(u_{\alpha},u_{\beta})^\top$ is its associated eigenvector, where the superscript $\top$ represents transpose and $u_{\alpha}$ and $u_{\beta}$ correspond to nodes in $\alpha$ and $\beta$, respectively.
Then, one can write the eigenvalue equation for matrix $P_{n+1}$ in the following block form:
\begin{equation}\label{T1}
\left[\begin{array}{cccc}
0 & P_{\alpha, \beta} \\
P_{\beta,\alpha} & 0
\end{array}
\right]
\left[\begin{array}{cccc}
 u_{\alpha} \\
 u_{\beta}
\end{array}
\right]={\lambda}_{i}(n+1) \left[\begin{array}{cccc}
 u_{\alpha} \\
 u_{\beta}
\end{array}
\right].
\end{equation}
Equation~(\ref{T1}) can be recast into two equations:
\begin{equation}\label{T3}
P_{\alpha, \beta}u_{\beta}=\lambda_{i}(n+1)u_{\alpha},
\end{equation}
\begin{equation}\label{T4}
P_{\beta,\alpha}u_{\alpha}=\lambda_{i}(n+1)u_{\beta}.
\end{equation}
Equation~(\ref{T4}) yields
\begin{equation}\label{T5}
u_{\beta}=\frac{1}{\lambda_{i}(n+1)}P_{\beta,\alpha}u_{\alpha}\,,
\end{equation}
provided that $\lambda_{i}(n+1)\neq 0$. Plugging Eq.~(\ref{T5})
into Eq.~(\ref{T3}) gives
\begin{equation}\label{T6}
\frac{1}{\lambda_{i}(n+1)}P_{\alpha,\beta}P_{\beta,\alpha}u_{\alpha}=\lambda_{i}(n+1)u_{\alpha}.
\end{equation}
Thus, we reduce the problem of determining the eigenvalues
for matrix $P_{n+1}$ of order $N_{n+1} \times N_{n+1}$
to calculating the eigenvalues of matrix $P_{\alpha,\beta}P_{\beta,\alpha}$ with a smaller
order of $N_{n} \times N_{n}$.

In the Appendix, we prove that
\begin{equation}\label{T7}
P_{\alpha,\beta}P_{\beta,\alpha}=\frac{2m+1}{2m+2}I_{n}+\frac{1}{2m+2}P_{n},
\end{equation}
where $I_{n}$ is the identity matrix of the same order as $P_{n}$. Thus, we have related $P_{\alpha,\beta}P_{\beta,\alpha}$ to $P_{n}$, which enables us to express the eigenvalues of matrix $P_{n+1}$ in terms of those of $P_{n}$.

Substituting Eq.~(\ref{T7}) into Eq.~(\ref{T6}), we obtain
\begin{equation}\label{T10}
\frac{1}{\lambda_{i}(n+1)}\left(\frac{2m+1}{2m+2}I_{n}+\frac{1}{2m+2}P_{n}\right)u_{\alpha}
=\lambda_{i}(n+1)u_{\alpha},
\end{equation}
that is,
\begin{equation}\label{T12}
P_n u_{\alpha}=\left\{(2m+2)[\lambda_{i}(n+1)]^2-(2m+1)\right\}u_{\alpha}.
\end{equation}
Hence, if $\lambda_{i}(n)$ is the eigenvalue of $P_{n}$ corresponding to the eigenvector $u_{\alpha}$, Eq.~(\ref{T12}) implies
\begin{equation}\label{T13}
\lambda_{i}(n)=(2m+2)[\lambda_{i}(n+1)]^2-(2m+1).
\end{equation}
Solving the quadratic equation in the variable $\lambda_{i}(n+1)$
given by Eq.~(\ref{T13}) yields
\begin{equation}\label{T16}
\lambda_{i,1}(n+1)= \sqrt{\frac{2m+1+\lambda_{n}}{2m+2}}, \quad \lambda_{i,2}(n+1)=-\lambda_{i,1}(n+1) \,.
\end{equation}
Equation~(\ref{T16}) relates $\lambda_{i}(n+1)$ to $\lambda_{i}(n)$, with each $\lambda_{i}(n)$ giving rise to two eigenvalues of $P_{n+1}$.

\subsection{Degeneracies of eigenvalues}

To determine the multiplicities of the eigenvalues, we first calculate numerically the eigenvalues for those networks of small sizes. For $F_0$, the eigenvalues are $1$ and $-1$; while for $F_1$, its  eigenvalues are $1$, $-1$, $0$, $\frac{\sqrt{m+m^2}}{m+1}$, and $\frac{-\sqrt{m+m^2}}{m+1}$. For $n\geq 2$, we find that the eigenvalue spectrum displays the following properties: (i) From Eq.~(\ref{T16}), eigenvalue $1$ gives rise to two eigenvalues $1$ and $-1$ with single degeneracy. (ii) All eigenvalues of a given generation $n_{i}$ always exist at its subsequent generation $n_{i}+1$, and all new eigenvalues at generation $n_{i}+1$ are just those generated via Eq.~(\ref{T16}) by substituting $\lambda_{i}(n)$ with $\lambda_{i}(n_i)$ that are newly added to generation $n_{i}$; moreover each new eigenvalue keeps the degeneracy of its father. (iii) Except for $0$, all other eigenvalues are generated from $-1$ and $0$, and all the offspring eigenvalues of $-1$ are nondegenerate. Thus, all that is left is to determine the degeneracy of $0$ and the multiplicities of its offsprings, based on property (ii).

Let $D^{\rm mul}_n(\lambda)$ denote the multiplicity of eigenvalue $\lambda$ of matrix $P_n$. We now find the number of eigenvalue 0 of $P_n$. Let $r(M)$ be the rank of matrix $M$. Then, the degeneracy of eigenvalue 0 of $P_{n+1}$ is
\begin{equation}\label{N0}
D^{\rm mul}_{n+1}(\lambda=0)= N_{n+1}-r(P_{n+1})\,.
\end{equation}
In order to determine $D^{\rm mul}_{n+1}(\lambda=0)$, we can alternatively compute $r(P_{n+1})$. Obviously, $r(P_{n+1})=r(P_{\alpha,\beta})+r(P_{\beta,\alpha})=2r(P_{\beta,\alpha})$, where $r(P_{\alpha,\beta})=r(P_{\beta,\alpha})$ is used.

We next determine $r(P_{\beta,\alpha})$. First, we show that $P_{\beta,\alpha}$ is a full column rank matrix. Let
\begin{equation}
v=(v_1,v_{2},\ldots,v_{N_{n+1}-N_{n}})^\top=\sum_{\substack{i = 1 \\ i \in \alpha }}^{N_{n}} k_i M_{i},
\end{equation}
where $M_{i}$ is the column vector of $P_{\beta,\alpha}$ as defined in Eq.~(\ref{App1}). Let $M_{i}=(M_{1,i},M_{2,i},\ldots,M_{N_{n+1}-N_{n},i})^\top$.
Suppose that $v=0$. Then, we can prove that for an arbitrary $k_i$, $k_i=0$ always holds.
By construction, for any old node $i \in \alpha$, there exists a new leaf node $l \in \beta$ attached to $i$. Then, for $v_l=k_1M_{1,l}+k_2M_{2,l},\ldots,k_{N_{n+1}-N_{n},l}M_{N_{n+1}-N_{n},l}$, only $M_{i,l}\neq0$ but all $M_{x,l}=0$ for $x\neq i$. From $v_l=0$, we have $k_i=0$. Therefore, $r(P_{\beta,\alpha})=N_n=(2m+2)^{n}+1$.

Combining the above-obtained results, we obtain the degeneracy of
eigenvalue 0 of $P_n$ as
\begin{equation}
D^{\rm mul}_n(\lambda=0)=\begin{cases}
0, &n=0, \\
2m(2m+2)^{n-1}-1, &n \geq 1.
\end{cases}
\end{equation}
Since every eigenvalue in $P_n$ keeps the degeneracy of its father, the multiplicity of each first-generation descendants of eigenvalue 0 is $2m(2m+2)^{n-2}-1$, the multiplicity of each second-generation descendants of eigenvalue 0 is $2m(2m+2)^{n-3}-1$, and so on. Thus, the total number of eigenvalue 0 and all of its descendants in $P_n$ ($n\geq 1$) is
 \begin{eqnarray}\label{N6}
N_n^{\rm seed}(0)&=&\sum_{i=1}^{n}
[2m(2m+2)^{i-1}-1]2^{n-i}\nonumber \\
&=&(2m+2)^{n}-2^{n+1}+1.
\end{eqnarray}
Similarly, the total number of eigenvalue $-1$ and its descendants in $P_n$ ($n\geq 0$) is
\begin{equation}\label{N6}
N_n^{\rm seed}(-1)=\sum_{i=0}^{n}2^{i}=2^{n+1}-1.
\end{equation}
Summing up the number of eigenvalues found above, we have
\begin{eqnarray}\label{N6}
N_n^{\rm seed}(0)+N_n^{\rm seed}(-1)+1=(2m+2)^{n}+1=N_n,
\end{eqnarray}
which means that we have determined all the eigenvalues of $P_n$.

\section{Application of eigenvalue spectrum} \label{app}

We next show how to use the obtained eigenvalues
and their multiplicities to determine some related quantities for the
fractal scale-free network family $F_n$, including the eigentime identity
of random walks and the number of spanning trees. We note that the number of spanning trees for every generation is 1, so our aim for evaluating spanning trees is to verify that our computation of the eigenvalues of the transition matrix is right.

\subsection{Eigentime identity}


Eigentime identity of random walks on a network is a global characteristic of the network, which reflects the structure of the whole network.

Let $H_{ij}(n)$ denote the mean-first
passage time from node $i$ to node $j$ in $F_{n}$,
which is the expected time for a walker starting from node $i$ to reach node $j$ for the first time. Let $\pi=(\pi_1, \pi_2,\ldots, \pi_N)^\top$ represent the stationary distribution for random walks on $F_{n}$~\cite{Lo96,AlFi99}. It is easy to derive $\pi_i=d_i(n)/(2E_n)$, satisfying  $\sum_{i=1}^{N}\pi_i=1$ and $\pi^{\top}T_n=\pi^{\top}$.
Then, the eigentime identity, denoted by $H_n$, for random walks on $F_{n}$, is the expected time from a node $i$ to another node $j$, selected randomly from all nodes
accordingly to the stationary distribution. That is,
\begin{equation}\label{eig01}
H_n=\sum_{j=1}^{N_{n}}\pi_j\,H_{ij}(n)=\sum_{i=2}^{N_{n}}\frac{1}{1-\lambda_{i}(n)},
\end{equation}
where we have assumed that $\lambda_{1}(n)=1$.

Note that for any $i$, $\sigma_{i}(n)=1-\lambda_{i}(n)$ is an eigenvalue of the normalized Laplacian matrix of $F_n$, defined as $L_n=I_n-P_n$~\cite{Ch97,ChLuVu03,ChZh07}. The normalized Laplacian matrix has found numerous applications~\cite{DoGoMeSa03,VoBl07,HwYuLeKa10}.
By definition,
\begin{equation}\label{eig02}
H_n=\sum_{i=2}^{N_{n}}\frac{1}{\sigma_{i}(n)}.
\end{equation}
The goal next is to explicitly evaluate this sum.

Let $\Omega_n$ be the set of all the $N_n-1$ nonzero eigenvalues of matrix
$L_n$, $\Omega_n=\{ \sigma_2(n),
\sigma_3(n),\ldots,\sigma_{N_n}(n)\}$, in which the
distinctness of the elements has been ignored. It is clear that $\Omega_n$ ($n\geq 1$) includes
$1$, $2$, and other eigenvalues generated by them. Then, $\Omega_n$ can be classified into three subsets represented by $\Omega_n^{(1)}$, $\Omega_n^{(2)}$ and $\Omega_n^{(3)}$, respectively. That is, $\Omega_n=\Omega_n^{(1)} \cup \Omega_n^{(2)} \cup \Omega_n^{(3)}$, where $\Omega_n^{(1)}$ consists of eigenvalue 1 with multiplicity $2m(2m+2)^{n-1}-1$,
$\Omega_n^{(2)}$ contains only eigenvalue 2 with a single degeneracy, and
$\Omega_n^{(3)}$ includes those eigenvalues generated by 1 and 2. Obviously, $\sum_{i \in \Omega_n^{(1)}}\frac{1}{\sigma_{i}(n)}=2m(2m+2)^{n-1}-1$ and $\sum_{i \in \Omega_n^{(2)}}\frac{1}{\sigma_{i}(n)}=\frac{1}{2}$.

Each eigenvalue $\sigma_i(n-1)$ in $\Omega_{n-1}^{(3)}$ generates two eigenvalues,  $\sigma_{i,1}(n)$ and $\sigma_{i,2}(n)$, belonging to $\Omega_{n}^{(3)}$, via the following equation:
\begin{equation}\label{eig03}
[\sigma_i(n)]^2 - 2\sigma_i(n) + \frac{\sigma_i(n-1)}{2(m+1)} = 0\,,
\end{equation}
which is easily obtained from Eq.~(\ref{T13}). According to Vieta's formulas, we have $\sigma_{i,1}(n)+ \sigma_{i,2}(n)= 2$ and $\sigma_{i,1}(n)\times \sigma_{i,2}(n) = \frac{\sigma_i(n-1)}{2(m+1)}$.
Then
\begin{equation}\label{eig04}
\frac{1}{\sigma_{i,1}(n)} + \frac{1}{\sigma_{i,1}(n)} = \frac{4(m+1)}{\sigma_i(n-1)},
\end{equation}
which implies that
\begin{equation}\label{eig05}
\sum_{\sigma_i(n) \in \Omega_n^{(3)}}\frac{1}{\sigma_{i}(n)}= 4(m+1)\sum_{\sigma_i(n-1) \in \Omega_{n-1}^{(3)}} \frac{1}{\sigma_{i}(n-1)}.
\end{equation}
Thus, we have
\begin{equation}\label{eig06}
H_n = 4(m+1)H_{n - 1} +[2m(2m+2)^{n-1}-1]+ \frac{1}{2}.
\end{equation}
Considering the initial condition $H_1=\frac{8m+3}{2}$, Eq.~(\ref{eig06}) can be solved by induction to yield
\begin{eqnarray}\label{eig07}
H_n&= &\frac{(6m^2+6m+1)(4m+4)^n}{(m+1)(4m+ 3)}\nonumber \\
&\quad&-2m(2m+2)^{n-1}+\frac{1}{8m+6}\,,
\end{eqnarray}
which can be expressed in terms of network size $N_{n}$ as
\begin{eqnarray}\label{eig08}
H_n&= &\frac{6m^2+6m+1}{(m+1)(4m+ 3)}(N_{n}-1)^{1+\ln 2/ \ln (2m+2)}\nonumber \\
&\quad&-\frac{m}{m+1}(N_{n}-1)+\frac{1}{8m+6}\,.
\end{eqnarray}
Thus, for large networks, i.e., $N_g\rightarrow \infty$,
\begin{eqnarray}\label{eig09}
 H_{n}\sim (N_{n})^{1+\ln2/\ln(2m+2)}=
(N_{n})^{1+1/f_B}\,,
\end{eqnarray}
increasing as a power-law function of the network size $N_{n}$ with the power exponent greater than 1. Note that the power-law exponent in Eq.~(\ref{eig09})  is consistent with the
general scaling for the average global MFPT (i.e., eigentime identity) previously obtained in~\cite{TeBeVo09}, but is comparable to the diverse behaviors of MFPT observed in real-life systems,  the reason for which was explained  in~\cite{CaBeIvCa12}.

\subsection{Spanning trees}

An important task in the study of networks is to determine the number of spanning
trees on different
networks~\cite{ZhLiWuZh10,ZhLiWuZo11,LiWuZhCh11,WuZhCh12,ZhWuLi12}.
A spanning tree of a connected graph  is defined as a maximal set of edges of the graph that contains no cycle, or as a minimal set of edges that connect all nodes.
It has been shown~\cite{Ch97,ChZh07} that for a connected network $G$ with $N$
nodes, the number of its spanning trees, $N_{\rm st}(G)$, is related
to the $N-1$ nonzero eigenvalues of its normalized Laplacian
matrix. According to~\cite{Ch97,ChZh07}, the number of spanning trees $N_{\rm st}(F_{n})$ for $F_n$ is determined by
\begin{equation}\label{ST01}
N_{\rm st}(F_{n})=\prod_{i=1}^{N_{n}}
d_i(n)\prod_{i=2}^{N_{n}}\sigma_i(n) \Bigg /
\sum_{i=1}^{N_{n}}d_i(n).
\end{equation}

We now determine the terms in the sum and in the products of Eq.~(\ref{ST01}).
It is evident that
\begin{equation}\label{ST02}
\sum_{i=1}^{N_{n}}d_i(n)=2E_n=(2m+2)^{n}.
\end{equation}
Let $\Delta_n$ and $\Lambda_n$ denote $\prod_{i=1}^{N_{n}}
d_i(n)$ and $\prod_{i=2}^{N_{n}}\sigma_i(n)$, respectively.
According to the above-obtained results, it is easy to show that they obey the following recursive relations:
\begin{equation}\label{ST03}
\Delta_n = \Delta_{n-1}\times (m+1)^{N_{n-1}}\times 2^{E_{n-1}}
\end{equation}
and
\begin{equation}\label{ST04}
\Lambda_n =2\left(\frac{1}{2m+2}\right)^{N_{n-1}-1}\Lambda_{n-1}.
\end{equation}
Using $\Delta_1=1$ and $\Lambda_1=2$, we can solve Eqs.~(\ref{ST03}) and~(\ref{ST04}) to obtain
\begin{equation}\label{ST05}
\Delta_n = (m+1)^n(2m+2)^{\frac{(2m+2)^n-1}{2m+1}}
\end{equation}
and
\begin{equation}\label{ST06}
\Lambda_n =2^{n+1}\left(\frac{1}{2m+2}\right)^{\frac{(2m+2)^n-1}{2m+1}}.
\end{equation}
Inserting Eqs.~(\ref{ST02}),~(\ref{ST05}) and~(\ref{ST06}) into Eq.~(\ref{ST01}) gives
$N_{\rm st}(F_{n})=1$, which proves that the technique and process of our computation on the eigenvalues and their degeneracies for the transition matrix of the network family $F_n$ are indeed correct.

\section{Conclusions}

Rich information about the structure and random-walk dynamics of a network can be extracted from the spectrum of the transition matrix of the network. In this paper, we have studied the transition matrix of a class of networks with scale-free and fractal behaviors that are observed in various real-world networked systems. The iterative construction of the networks allows a detailed analysis of the eigenvalues and their multiplicities. Using the renormalization approach, we have derived an explicit recursive relationship between the eigenvalues of the network family at two successive iterations. For each eigenvalue, we have also determined its degeneracy.  Based on the obtained recursion relation of eigenvalues, we have further determined the eigentime identity for random walks, an important quantity rooted in the inherent topology of a network. Moreover, we have tested our computation and results for the spectrum of the transition matrix via evaluating the number of spanning trees in the network family under consideration, using the established relation between the two quantities. Our work opens up the possibilities to characterize the spectrum for the transition matrices of other deterministic networks.

It deserves to be mentioned that by definition eigentime identity is the expected time from a node $i$ to another node $j$ that is chosen randomly from all nodes according to the stationary distribution, which means that the target is immobile. Then, interesting questions arise: What happens if the target is not immobile? Would it be possible in this case to use the so-called Pascal Principle~\cite{TeBeVo09,MoOsBeCo03,MoOsBeCo04} to obtain a rigorous upper bound? Although the questions go beyond the scope of the present paper, they are worth studying in the future. Futhermore, future work should also include how to apply the theory to real-world networks, a relation between whose topology and function has been demonstrated~\cite{BaBaKaHaIv12}.

\begin{acknowledgments}
We would like to thank Bin Wu for assistance. This work was supported by the National Natural Science Foundation
of China under Grant No. 61074119 and the Hong Kong Research Grants
Council under the GRF Grant CityU 1114/11E.
\end{acknowledgments}


\section{Proof of equation~(\ref{T7})\label{AppA}}

In order to prove the equivalent relation $P_{\alpha,\beta}P_{\beta,\alpha}=\frac{2m+1}{2m+2}I_{n}+\frac{1}{2m+2}P_{n}$, it suffices to show that the entries of the former are equal to their counterparts of the latter. For matrix $\frac{2m+1}{2m+2}I_{n}+\frac{1}{2m+2}P_{n}$, denoted by $Q_n$, obviously its entries are:  $Q_n(i,i)=\frac{2m+1}{2m+2}$ for $i\neq j$ and $Q_n(i,j)=\frac{1}{2m+2}P_n(i,j)$ otherwise.

We proceed to determine the entries of $P_{\alpha,\beta}P_{\beta,\alpha}$, denoted by $R_n$. Note that $P_{\alpha,\beta}$ can be written as
\begin{equation} \label{App1}
P_{\alpha,\beta} =
\left(
\begin{array}{c}
M_1^\top\\
M_2^\top\\
\vdots\\
M^\top_{N_{n}}
\end{array}
\right),
\end{equation}
where each $M_i$ is a column vector with order $N_{n+1}-N_{n}$. Since $P_{\alpha,\beta}=P_{\beta,\alpha}^\top$, we have $P_{\beta,\alpha}=\left(
M_1~M_2~ \ldots M_{N_{n}}\right)$. Then, the entry $R_n(i,j)=M_i^\top M_j$ of $R_n$ can be determined in the following way.

If $i=j$, the diagonal element of $R_n$ is
\begin{eqnarray}\label{App2}
&\quad&R_n(i,i)= M_i^\top M_i \nonumber \\
&=& \displaystyle \sum_{k \in \beta} P_{n+1}(i,k)P_{n+1}(k,i)=\displaystyle \sum_{k \in \beta} \frac{A_{n+1}(i,k)}{d_i(n+1)d_k(n+1)} \nonumber \\
&=& \frac{1}{d_i(n+1)}\left(\sum_{\substack{k \in \beta, i\thicksim k \\ d_k(n+1) = 2 }} \frac{1}{2} + \sum_{\substack{k \in \beta, i\thicksim k \\ d_k(n+1) = 1}}1\right) \nonumber \\
&=& \frac{1}{d_i(n+1)}\left[d_i(n) \times\frac{1}{2} + m\,d_i(n) \right]\nonumber \\
&=& \frac{2m+1}{2m + 2}=Q_n(i,i),
\end{eqnarray}
where the relation $d_i(n+1)=(m+1)\,d_i(n)$ has been used. In Eq.~(\ref{App2}), $i\thicksim k$
means that nodes $i$ and $k$ are adjacent in $F_{n+1}$.

If $i \neq j$, the non-diagonal element of $R_n$ is
\begin{eqnarray}\label{App3}
&\quad&R_n(i,j)= M_i^\top M_j \nonumber \\
&=&  \sum_{k \in \beta} P_{n+1}(i,k)P_{n+1}(k,j) \nonumber \\
&=& \sum_{\substack{A_{n+1}(i,k) = 1 \\ A_{n+1}(k,j) = 1 }} \frac{A_{n}(i,j)}{d_k(n+1)\sqrt{d_i(n+1)d_j(n+1)}} \nonumber \\
&=& \frac{A_{n}(i,j)}{(2m+1)\sqrt{d_i(n)d_j(n)}}=Q_n(i,j).
\end{eqnarray}
Thus, Eq.~(\ref{T7}) is proved.

\end{document}